\documentclass{article}
\renewcommand{\(}{\left(}
\renewcommand{\)}{\right)}
\newcommand{\beq}{\begin{equation}}
\newcommand{\eeq}{\end{equation}}
\newcommand{\beqa}{\begin{eqnarray}}
\newcommand{\eeqa}{\end{eqnarray}}

\newcommand{\bin}[2]{\left( \begin{array}{c} #1 \\ #2 \end{array} \right) }
\newcommand{\arr}{{\em arr}}
\newcommand{\dep}{{\em dep}}
\newcommand{\w}{\omega}
\newcommand{\half}{\frac{1}{2}}
\newcommand{\eq}[1]{eq. (\ref{#1})}
\newcommand{\fig}[1]{fig. \ref{#1}}
\usepackage{graphicx}

\begin{document}
\begin{titlepage}
\begin{flushright}
LU TP 97-11 \\
Revised \\
\end{flushright}
\vspace{0.8in}
\LARGE
\begin{center}

{\bf Statistical Properties of Unrestricted \\ Crew Scheduling Problems} \\
\vspace{0.3in}
\large
Martin Lagerholm\footnote{martin@thep.lu.se},
Carsten Peterson\footnote{carsten@thep.lu.se} and 
Bo S\"{o}derberg\footnote{bs@thep.lu.se}\\
\vspace{0.1in}
Computational Biology and Biological Physics\\ 
University of Lund, S\"{o}lvegatan 14A, S-223 62 Lund, Sweden\\
\vspace{0.3in}


\end{center}
\vspace{0.2in}
\normalsize

Abstract:

A statistical analysis is performed for a random unrestricted local
crew scheduling problem, expressed in terms of pairing arrivals with
departures.  The analysis is aimed at understanding the structure of 
similar problems with global restrictions, and estimating their
difficulty. The methods developed are of a general nature and can be
of use in other problems with a similar structure. For large random
problems, the ground-state energy scales like $\sqrt{N}$ and the
average excitation like $N$, where $N$ is the number of
arrivals/departures. The average ground-state degeneracy is such that
the probability of hitting an optimal pairing by chance scales like
$2N2^{-N}$ for large $N$. By insisting on the local ground-state
energy for a restricted problem, airports can be split into smaller
parts, and the state space reduced by typically a factor $\sim 2^{N_a}$,
with $N_a$ the total number of airports.

\end{titlepage}


\newpage

%
\section{Introduction}
%

Airline crew scheduling represents an important class of optimization
problems where the topological structure is important.  In
refs. \cite{Potts,long}, a novel Potts artificial neural network approach is
developed for attacking semi-realistic airline crew scheduling
problems of the following type:
A weekly flight schedule is given in terms of a set of flights, each
with a specified airport and time for departure and arrival. The
object is to assign a crew to each flight, while minimizing a
cost-function defined by the total required crew time (including the
waiting-time at airports). The solution is subject to a set of
global constraints: The crews are required to travel along closed
tours, starting and ending at a certain airport, the home-base. These
tours are subject to limitations as to duration and leg-count.

As shown in ref. \cite{long}, a great deal of simplification is gained
by reformulating the problem as that of mapping arrivals on departures 
at each airport, implying an implicit representation of the crews.  In
fact, without the global restrictions, the problem is reduced to a set
of independent local subproblems, one at each airport. Each local
problem amounts to minimizing the local waiting-time, and is simply
solvable in polynomial time.

In this paper, we focus on this kind of unrestricted local problems,
in particular their statistical properties. These are quite
interesting, and in no way trivial, in spite of the triviality of the
problems. In particular, we consider the ensemble of {\em random}
local problems of a fixed size $N$, as defined by the number of
arrivals/departures. In addition, we analyze the properties of random
solutions to such problems.

Such a statistical analysis of this type of problem does not exist in
the literature, and we feel it is interesting for the following
reasons: The results illuminate the structure of the corresponding
restricted problems, and as a by-product, useful tools are provided
for probing their difficulty, and for simplifying their solution.
Some of the methods used are novel, and may be used also in other
contexts, where a similar structure occurs.  In addition, a lower
bound to the waiting-time is provided by the solutions to the 
unrestricted problem. This bound is often saturated \cite{long}.

The methodology we use contains the following steps.  First, the
analysis of a local problem is simplified by considering its topology
(defined by the relative ordering in time between arrivals and
departures) separately from its geometry (defined by the lengths of
the time intervals between consecutive events).  The ensemble of
problems is thus factorized into the direct product of the {\em
ensemble of topologies} and the {\em ensemble of geometries}.

After introducing a notation for the topology, we consider problems
with a fixed topology, and evaluate averages over the geometry of
entities related to the waiting-time. These are simple, since the effect
of the geometry on the waiting-time spectrum is a mere shift.

The apparent difficulty of a problem is probed by analyzing the
distribution of waiting-times of random solutions. A nice feature is that
the waiting-time spectrum for each problem is quantized in steps of the
basic period of the schedule.

Subsequently, all variables of interest are averaged also over the
topology. This is a more difficult task, and requires the use of a
subtle recursive method.

An alternative measure of the difficulty of a problem is the
ground-state degeneracy, i.e. the number of solutions with minimal
waiting-time, as compared to the total number of solutions, given by
$N!$. This is independent of geometry. Due to the character of the
dependence on topology, non-standard methods are required to compute
the average over topology.

This paper is organized as follows: In Section 2, the problem ensemble
is defined.  In Section 3, a formalism is introduced for
characterizing the topology. The degeneracy structure as a function of
topology is analyzed, and various energy moments for fixed topology
are computed, by averaging over geometry and/or pairing.  In Section 
4, a statistical analysis of the detailed degeneracy structure is
considered. In particular, the average ground-state degeneracy is
computed.  Section 5 contains our conclusions.

%
\section{Unrestricted Crew Scheduling}
%

%
\subsection{The Local Problem}
%

A local problem of size $N$ is defined by specifying the times for $N$
arrivals (\arr's) and $N$ departures (\dep's), denoted respectively by
$t^a_i$ and $t^d_i, \; i = 1 \ldots N$.  The object is simply to find
a one-to-one mapping (a {\em pairing}) between the \arr's and \dep's,
such that the {\em energy} (or objective function) $E$, given by the
total waiting-time, is minimal. In general this can be done in more than
one way, implying a degeneracy of the ground-state.

The pairing of an \arr\ $A$ with a \dep\ $D$ implies that the crew
assigned to $A$ should next be assigned to $D$. The periodicity of the
schedule implies that any \arr\ may be mapped on any \dep: if the
\dep\ is earlier, it is taken as the same \dep\ in the next period.

In what follows, we will use the period as the unit of time (and thus
energy). Then the times for the \arr's and \dep's can be restricted to
the unit interval. If an \arr\ $i$ is mapped on a \dep\ $j$, their
contribution to the total waiting-time is given by
\beq
	t^w_{ij} = \( t^d_j - t^a_i \) \mbox{ mod } 1 \; \in [0,1].
\eeq
Thus, whatever the mapping, the energy $E$ is restricted to the
interval $[0,N]$, and between different pairings it can only change by
an integer amount. Pairings yielding the lowest possible energy $E_0$ 
are said to belong to the ground-state, while a pairing with
$E=E_0+k,\;k>0$ is said to belong to $k$:th excited state.
\begin{figure}[htb]
\centering
\includegraphics[width=14cm]{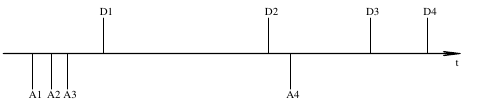}
\caption{An example of a local flight schedule.}
\label{sched}
\end{figure}
%
In figure \ref{sched}, an example of a flight schedule is depicted.

%
\subsection{The Random Local Problem}
%

A random local problem is defined by independently choosing the times
for $N$ \arr's and $N$ \dep's randomly on the unit period. It is
useful to divide the characterization of a given problem in two parts:
its {\bf topology} and its {\bf geometry}. The topology is defined by
the relative ordering in the combined (cyclic) sequence of \arr's and
\dep's. The probability is the same for every distinct topology. The
geometry is defined by the sizes of the $2N$ inter-spaces
$x_i,i=1,\ldots,2N$, into which the period is divided.

For a given problem, the solution space is defined by the set of {\bf
pairings}, i.e. the $N!$ possible mappings between \arr's and \dep's.

We will be interested mainly in the statistical properties of the
following entities: the ground-state energy (minimal waiting-time) $E_0$,
the energy $E$ of a random pairing, and the (integer) difference $D =
E - E_0$ defining the excitation energy. We will also be interested in
the degeneracies of the ground-state and the excited states.  To that
end, we will consider three kinds of averages: over respectively the
{\em topology}, the {\em geometry}, and the {\em pairing}.

The degeneracies of the ground-state and the excited states depend
only on the topology, i.e. the combined ordering of \arr's and \dep's.

For a fixed topology, the excitation energy $D$ is independent of the
geometry, and depends entirely on the pairing. Conversely, the
ground-state energy $E_0$ obviously is independent of the pairing, and
depends only on the geometry. Thus, for a fixed topology, $D$ and
$E_O$ are completely {\em uncorrelated}, in the combined ensemble of
random geometries and random pairings.

%
\section{Analysis for Fixed Topology}
%

%
\subsection{Characterization of the Topology}
%

\begin{figure}[htb]
\centering
\includegraphics[width=14cm]{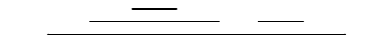}
\caption{Graphical representation of the topology $[AAADDADD]$}
\label{exlines}
\end{figure}

A simple way to achieve a ground-state pairing (i.e. solve the problem)
for a given topology is as follows:
\begin{enumerate}
\item An \arr\ immediately followed by a \dep\ is paired with that \dep,
	and both are removed from the sequence. Note that the \dep\
	could be in the next period.
\item The process is continued until all \arr's and \dep's are used.
\end{enumerate}
As an example, consider the sequence $[A_1 A_2 A_3 D_1 D_2 A_4 D_3
D_4]$, corresponding to the topology of the schedule in figure
\ref{sched}.
\begin{itemize}
\item Pairing $A_3$ with $D_1$ leaves $[A_1 A_2 D_2 A_4 D_3 D_4]$.
\item Pairing $A_2$ with $D_2$ leaves $[A_1 A_4 D_3 D_4]$.
\item Pairing $A_4$ with $D_3$ leaves $[A_1 D_4]$.
\item Pairing $A_1$ with $D_4$ finishes the process.
\end{itemize}
A graphical representation of the topology is now defined as follows.
\begin{itemize}
\item If necessary, rotate the sequence such that no pairing crosses
	the interval border.
\item Represent the \arr's and \dep's by equally spaced points on the
	interval.
\item For each pairing in turn, draw a line from the \arr\ to the
	\dep\ on the lowest level (one). All previously drawn lines
	that overlap with the new line are lifted one level.
\end{itemize}
Thus a set of lines at different levels are obtained, each line
starting at an \arr\ and ending on a \dep. For the example above, the
result is shown in \fig{exlines}.
Each line represents a crew waiting for a \dep.

It turns out that for the properties of the total energy spectrum,
full knowledge of the topology is not necessary; it suffices to know
how many lines there are at the different levels. Thus let $P_k$ be
the number of lines at level $k$ ($k = 1,2,3,\ldots$). For the example
above, the $P$ sequence is $[1,2,1]$, i.e. $P_1 = 1,P_2 = 2,P_3 = 1$,
while $P_k=0$ for $k>3$.

%
\subsection{Degeneracy Structure}
%

The ground-state degeneracy is now given by the product of the
line-levels, i.e.
\beq
	g_0 = \prod_{k\ge 1} k^{P_k}.
\eeq
This is because each \dep\ must terminate a line, and the number of 
available crews is equal to the number of lines alive at that point,
which is given by the line's level. For the example we get $3 \times 2
\times 2 \times 1 = 12$, corresponding to half of the 24 possible
pairings.

Naively, the degeneracy of the $m$th excited state can be obtained by
adding $m$ dummy lines, covering the entire interval; they represent
extra crews. Then each \dep\ has $m$ additional crews to choose
between, and we get for the naive degeneracy\footnote{For the example
above, this precisely corresponds to the spectrum ${\bf O} |n\!> = a_n
|n\!>$ of the quantum-mechanical operator ${\bf O} = {\bf a}{\bf
a}{\bf a}{\bf a}^\dagger {\bf a}^\dagger {\bf a}{\bf a}^\dagger {\bf
a}^\dagger $, where ${\bf a}^\dagger, {\bf a}$ are harmonic oscillator
creation and annihilation operators, satisfying $[{\bf a},{\bf
a}^\dagger] = 1$, and $|n\!>\propto\({\bf a}^\dagger\)^n|0\!>$ is the
$n$:th excited state.}
\beq
\label{a_m}
	a_m = \prod_{k\ge 1} (k+m)^{P_k} ,
\eeq
where we have assumed indistinguishable dummy lines; otherwise we
would have an extra factor $m!$. This defines an infinite sequence.

However, some of the possible pairings will contain permanently
grounded crews (closed lines), or lines extending over more than one
period. The contribution to the naive multiplicity from solutions with
$n$ grounded crews and/or excessive periods depends on the proper
degeneracy $n$ steps down. Denoting the {\bf proper degeneracies} by
$g_m$, the relation is
\beq
	a_m = \sum_{n=0}^{m} g_{m-n} \bin{N+n}{n},
\eeq
where the last factor is a binomial coefficient. This represents a
kind of renormalization, and can be inverted to yield the proper
degeneracies
\beq
\label{ga}
	g_m = \sum_{n=0}^{m} a_{m-n} (-)^n \bin{N+1}{n}.
\eeq
This must define a finite sequence, since $g_m \ge 0$ and the total
number of pairings, $\sum_m g_m = N!$, is finite.

From eq. (\ref{a_m}), it is obvious that the naive degeneracy $a_m$ is
an $N$th degree polynomial in $m$; hence it can be written as
\beq
\label{c_k}
	a_m = \sum_{k=0}^{N} c_k \bin{m+k}{k},
\eeq
with some coefficients $c_k$. Define the {\em generating functions}
\beq
	A(x) = \sum_m a_m x^m \;,\;\;
	G(x) = \sum_m g_m x^m.
\eeq
Due to eq. (\ref{ga}), these are related by $G(x) = (1-x)^{N+1}
A(x)$. Then, in terms of $c_k$, we have
\beqa
	A(x) &=& \sum_{k=0}^N c_k (1-x)^{-k-1},
\\
	G(x) &=& \sum_{k=0}^N c_k (1-x)^{N-k}.
\eeqa
From this, we see that the $g$ sequence is indeed finite: $g(m)=0$ for
all $m > N$. The individual degeneracies $g_m$ can be obtained from
$G$ and its derivatives at $x=0$,
\beqa
	g_0 &=& G(0) \; = \; \sum_k c_k,
\\ \nonumber
	g_1 &=& G'(0) \; = \; -\sum_{k=0}^N (N-k) c_k,
\eeqa
etc.

%
\subsection{The Excitation Energy for a Random Pairing}
%

Conversely, the moments over $g_m$ are readily obtained from $G$ and
its derivatives at $x=1$,
\beqa
	\sum_m g_m &=& G(1) \; = \; c_N,
\\ \nonumber
	\sum_m m g_m &=& G'(1) \; = \; -c_{N-1},
\\ \nonumber
	\sum_m m(m-1) g_m &=& G''(1) \; = \; 2 c_{N-2},
\eeqa
etc.  In order to relate $c_k$ to $P_k$, we can express the
coefficients of the polynomial $a_m$ in two different ways. From
eq. (\ref{a_m}) we get for the leading coefficients
\beq
	a_m
	= m^N
	+ m^{N-1} \sum_k k P_k
	+ \frac{m^{N-2}}{2} \left\{ \( \sum_k k P_k \)^2 - \sum_k k^2 P_k \right\}
	+ \ldots,
\eeq
while eq. (\ref{c_k}) gives, upon expanding the binomial coefficients,
\beqa
	a_m &=&
	\frac{m^N}{N!} c_N
	\; + \;
	\frac{m^{N-1}}{N!} \left\{ \half N(N+1) c_N + N c_{N-1} \right\}
\\ \nonumber
	&+&
	\frac{m^{N-2}}{N!} \left\{
	\frac{1}{24} N(N-1)(N+1)(3N+2) c_N + \half N^2(N-1) c_{N-1} + N(N-1) c_{N-2}
	\right\} + \ldots
\eeqa
From this we obtain the relations
\beqa	
	c_N &=& N!,
\\ \nonumber
	c_{N-1} &=& N! \left\{
	\frac{1}{N} \sum_k k P_k - \frac{N+1}{2}
	\right\},
\\ \nonumber
	c_{N-2} &=& N! \left\{
	\frac{1}{2N(N-1)} \(\sum_k k P_k\)^2
	- \frac{1}{2N(N-1)} \sum_k k^2 P_k
	- \half \sum_k k P_k
	+ \frac{(N+1)(3N-2)}{24} 
	\right\},
\eeqa
etc. This gives (as it should) $\sum_m g_m = N!$, the number of
possible pairings.  In a given topology, the fraction of pairings
having an energy $D=m$ steps above $E_0$ is given by $g_m / N!$. Thus,
the first few moments of the excitation energy $D$ for a random 
pairing are:
\beqa
\nonumber
	<D>_p &=& \frac{1}{N!} \sum_m m g_m
	\; = \; \frac{N+1}{2} - \frac{1}{N} \sum_k k P_k,
\\ \label{aveD}
	<D^2>_p &=& \frac{1}{N!} \sum_m m^2 g_m
\\ \nonumber
	&=& \frac{1}{N(N-1)} \(\sum_k k P_k\)^2
	- \frac{1}{N(N-1)} \sum_k k^2 P_k
	- \frac{N+1}{N} \sum_k k P_k
	+ \frac{(N+1)(3N+4)}{12},
\eeqa
where $<>_p$ denotes the average over pairings for a fixed topology.

%
\subsection{The Ground-State Energy in a Random Geometry}
%

For a given topology, the ground-state energy depends on the geometry,
and is given by
\beq
	E_0 = \sum_{i=1}^{2N} k_i x_i,
\eeq
where $x_i$ is the length of the $i$th sub-interval, and $k_i$ the
number of lines in that interval.

To perform averages over a random geometry, we need to analyze the
distribution of the intervals $x_i$. Independently of the topology,
they obey the distribution
\beq
	dP = (2N-1)! \; \delta \( \sum_i x_i - 1 \) \prod_{i=1}^{2N}
	\( \Theta(x_i) dx_i\).
\eeq
For a single $x_i$, this implies the distribution
\beq
	f(x_i) = (2N - 1) \(1 - x_i\)^{2N-2} \; , \; 0 \le x_i \le 1.
\eeq
Using the identity $\sum_i x_i = 1$, and the permutation symmetry
between different $x_i$, we have 
\beqa
	<x_i> &=& \frac{1}{2N},
\\
	<x_i x_j> &=& \frac {1 + \delta_{ij}} {2N(2N+1)}.
\eeqa

In what follows, we will also need the number of intervals with $k$
lines, to be denoted by $Q_k$; in terms of $P_k$ (defining $P_0 = 0$) it is simply
\beq
	Q_k = P_k + P_{k+1}, \; k \ge 0,
\eeq
since adding a line on level $k$ implies adding two intervals, with
respectively $k$ and $k-1$ lines. Obviously, we have $\sum_k Q_k =
2N$.

Now we are ready to compute the first few moments of $E_0$, yielding
\beqa
\nonumber
	<E_0>_g &=& \sum_{i=1}^{2N} k_i <x_i>_g = \frac{1}{2N} \sum_i k_i
	= \frac{1}{2N} \sum_k k Q_k
	= \frac{1}{N} \sum_k k P_k - \half,
\\ \nonumber
	<E_0^2>_g &=& \sum_{ij} k_i k_j <x_i x_j>
	= \frac{1}{2N(2N+1)} \sum_{ij} k_i k_j (1 + \delta_{ij})
\\ \label{aveE0}
	& = & \frac{1}{2N(2N+1)} \( \sum_{kl} k l Q_k Q_l + \sum_k k^2 Q_k \)
\\ \nonumber
	& = & \frac{1}{2N(2N+1)} \(
	4 \(\sum_{k} k P_k\)^2
	+ 2 \sum_k k^2 P_k
	- 2 (2 N + 1) \sum_k k P_k
	+ N(N+1)
	\),
\eeqa
where $<>_g$ denotes average over the geometry for fixed topology.

\subsection{The Total Energy}
%

The total energy $E$ is given by the ground-state and excitation
energies, $E = E_0 + D$.  Averaging over both geometry and pairing, we
obtain the surprisingly simple result
\beq
	<E>_{gp} \,=\, <E_0>_g + <D>_p \,=\, \frac{N}{2},
\eeq
independently of topology.

Since we already have the second moments of $E_0$ and $D$, we only
need $<E_0D>$ to compute the second moment of $E$. Using the fact that
$E_0$ and $D$ are statistically independent for a fixed topology, we
get
\beq
\label{E0D}
	<E_0D>_{gp} \,=\,
	<E_0>_g<D>_p \,=\,
	- \frac{1}{N^2} \( \sum_k k P_k \)^2
	+ \frac{N+2}{2N} \sum_k k P_k
	- \frac{N+1}{4}.
\eeq
This implies
\beq
\label{aveE}
	<E^2>_{gp} =
	\frac {2(N^2+N+1)} {N^2(N-1)(2N+1)} \(\sum_k k P_k\)^2
	- \frac{3(N+1)}{N(N-1)(2N+1)} \sum_k k^2 P_k
	+ \frac{(N+1)(6N^2-N+4)} {12(2N+1)}.
\eeq
%

%
\section{Topology Statistics}
%

%
\subsection{The Topology Ensemble}
%

We now want to perform averages also over the topology, for the
moments of the ground-state energy $E_0$, the excitation energy $D$,
and the total energy $E$.
\begin{table}[b]
\centering
\begin{tabular}{|c|c|c|l|l|l|l|c|c|}
\hline
%
%
$N$ & Pattern & Prob. & $\{P_k\}$ & $\{Q_k\}$ & $\{a_i\}$ & $\{g_i\}$ & $\langle E_0\rangle$ & $\langle D\rangle$ \\
\hline
\hline
1 &	AD &		1 &	[1] &		[1,1] &		1,2,3,\ldots &		(1)	& 1/2 & 0 \\
\hline												            
2 &	AADD &		2/3 &	[1,1] &		[1,2,1] &	2,6,12,\ldots &		(2)	& 2/2 & 0 \\
2 &	ADAD &		1/3 &	[2] &		[2,2] &		1,4,9,\ldots &		(1,1)	& 1/2 & 1/2 \\
\hline												            
3 &	AAADDD &	3/10 &	[1,1,1] &	[1,2,2,1] &	6,24,60,\ldots &	(6)	& 9/6 & 0 \\
3 &	AADADD &	3/10 &	[1,2] &		[1,3,2] &	4,18,48,\ldots &	(4,2)	& 7/6 & 1/3 \\
3 &	AADDAD &	3/10 &	[2,1] &		[2,3,1] &	2,12,36,\ldots &	(2,4)	& 5/6 & 2/3 \\
3 &	ADADAD &	1/10 &	[3] &		[3,3] &		1,8,27,\ldots &		(1,4,1) & 3/6 & 3/3 \\
\hline												            
4 &	AAAADDDD &	4/35 &	[1,1,1,1] &	[1,2,2,2,1] &	24,120,360,\ldots &	(24)	& 8/4 & 0 \\
4 &	AAADADDD &	4/35 &	[1,1,2] &	[1,2,3,2] &	18,96,300,\ldots &	(18,6)	& 7/4 & 1/4 \\
\parbox{2mm}{4 4} & \parbox{21mm}{AAADDADD AADAADDD} & $\left. \parbox{7mm}{4/35 4/35}\right\}$ & [1,2,1] & [1,3,3,1] & 12,72,240,\ldots & (12,12) & 6/4 & 2/4 \\
4 &	AAADDDAD &	4/35 &	[2,1,1] &	[2,3,2,1] &	6,48,180,\ldots &	(6,18)	& 5/4 & 3/4 \\
4 &	AADADADD &	4/35 &	[1,3] &		[1,4,3] &	8,54,192,\ldots &	(8,14,2) & 5/4 & 3/4 \\
\parbox{2mm}{4 4} & \parbox{21.1mm}{AADDAADD AADADDAD} & $\left. \parbox{7mm}{2/35 4/35}\right\}$ & [2,2] & [2,4,2] & 4,36,144,\ldots & (4,16,4) & 4/4 & 4/4 \\
4 &	AADDADAD &	4/35 &	[3,1] &		[3,4.1] &	2,24,108,\ldots &	(2,14,8) & 3/4 & 5/4 \\
4 &	ADADADAD &	1/35 &	[4] &		[4,4] &		1,16,81,256,\ldots & (1,11,11,1) & 2/4 & 6/4 \\
\hline
\end{tabular}
\caption{Inequivalent topologies for $N\le 4$. For each topology, the
third column gives the probability of occurrence, while the next two
give its $P_k$ and $Q_k$ sequence, respectively. In the following two
columns the $a_i$ and $g_i$ sequences are given, whereas the last two
columns give the average ground-state energy and excitation,
respectively. Note that different topologies might have identical
characteristics in terms of $P_k$, etc.}
\label{topol}
\end{table}
In table \ref{topol} a complete list of topologies for $N$ up to four
is given, along with various characteristics. Note the similarity in
symmetry-properties between the $P$, $Q$ and $g$ sequences.

To compute $<P_k>$, we must analyze the number of ways $m[P]$ to
obtain a given $P$ sequence.
\begin{itemize}
\item The lowest level lines define $P_1$ non-overlapping subgroups;
	these are cyclically indistinguishable, so the naive
	multiplicity must be divided by $P_1$.
\item For the $P_k$ indistinguishable lines on a higher level $k$, there
	are $P_{k-1}$ possible lines on the previous level to put them
	above. This can be done in $\bin{P_k + P_{k-1} - 1}{P_k}$
	ways.
\item In addition, there are $2N$ cyclic rotations of the $AD$
	sequence that yield the same topology.
\item The $P$ sequence must sum up to $N$, and $P_1$ must not vanish.
\end{itemize}
This gives for the multiplicity $m[P]$
\beq
	m[P] = \frac{2N}{P_1} \delta_{N - \sum P_k} \prod_{k=1}^{\infty}
	\bin {P_k+P_{k+1}-1} {P_{k+1}}.
\eeq

The total number of possible topologies for a fixed $N$ is simply the
number of possible $AD$ sequences, i.e. the number of distinct
orderings of a sequence of $N$ $A$'s and $N$ $D$'s. This is given by
$\bin{2N}{N}$.  This should match the total multiplicity $M_N \equiv
\sum_{[P]} m([P])$, which can be recursively obtained from $m[P]$ by
expanding the Kronecker delta in terms of a complex integral along a
{\em small} contour $C$ around the origin:
\beq
	\delta_{N-\sum P_k} = \oint_C \frac{dz}{2 \pi i z^{N+1}} \prod z^{P_k}.
\eeq
By iteratively using the identity
\beq
\label{binsum}
	z^a \sum_{b=0}^{\infty} \bin{a+b-1}{a - 1} w^b
	= z^a \sum_{b=0}^{\infty} \bin{-a}{b} (-w)^b = \(\frac{z}{1-w}\)^a.
\eeq
when summing $m[P] \prod z^{P_k}$ over the $P_k$ in order of
decreasing $k$, we obtain convergence of the power factor to
$\w^{P_k}$, with $\w$ given by
\beq
	\w = \frac{z}{1 - \w} \; \Rightarrow \; \w = \half \(1 - \sqrt{1-4z} \).
\eeq
Thus, the total multiplicity is given by
\beq
\label{mult}
	M_N =
	2N \oint_C \frac {dz\; z^{-N}} {2\pi i z} \sum_{P_1} \frac{\w(z)^{P_1}}{P_1}
	=
	2N \oint_C \frac {dz\; z^{-N}} {2\pi i z} \( -log(1-\w(z)) \)
	=
	\bin{2N}{N}, \eeq
obtained by extracting the $z^N$ coefficient of $-\log(1-\w(z))$ as
$\frac{1}{2N}\bin{2N}{N}$, by doing the integral in terms of $w$ using
$z=w(1-w)$.

%
\subsection{Distribution of $P_1$}
%

The number of non-overlapping groups of lines is given by the number
of lines at level 1, i.e. $P_1$.  In fact, $w(z)$ can be seen as a
generating function,
\beq
	w(z) = \sum_{n=1}^{\infty} m^{(1)}_n z^n
	= -\half \sum_{n=1}^{\infty} (4z)^n \bin{n-\frac{3}{2}}{n}
	= \half \sum_{n=1}^{\infty} (2z)^n \frac{(2n-3)!!}{n!},
\eeq
for the number of ways ($m^{(1)}_n$) to arrange $n$ lines in such a
group. The distribution of $P_1$ is easy to compute by slightly
modifying eq. (\ref{mult}): skipping the summation over $P_1$ and
dividing by $M_N$ yields
\beq
	\mbox{Prob}(P_1) =
	\frac{2N}{\bin{2N}{N}} \oint_C \frac {dz\; z^{-N}} {2\pi i z} \frac{\w(z)^{P_1}}{P_1}
	= \frac{N!(2N-1-P_1)!}{(2N-1)!(N-P_1)!}
\eeq
for $P_1>0$. For large $P_1 \ll N$, this is close to $2^{-P_1}$.

The grouping of lines corresponds to a grouping of the flights, with a
matching number of \arr's and \dep's in each group. In a ground-state
pairing, flights in different groups are never paired, and within a
group, the \arr\ has to precede its paired \dep. This fact can be used
to simplify also restricted problems.

%
%

%
\subsection{Moments of $P_k$}
%

Similarly, $<P_k>$ can be obtained from realizing that inserting a
factor $P_k$ in the sum over $P_k$ gives a factor
$P_{k-1}\w/(1-\w)$. This yields in the end an extra factor
$P_1(\w/(1-\w))^{k-1}$ in the $P_1$ sum:
\beqa
	M_N <P_k>
	&=& 2N \oint_C \frac{dz}{2\pi i z^{N+1}} \(\frac{\w}{1-\w}\)^{k-1} \sum_{P_1>1} \w^{P_1}
\\
	&=& 2N \oint_C \frac{dz}{2\pi i z^{N+1}} \(\frac{\w}{1-\w}\)^k
\\
	&=& 2N \oint_C \frac{d\w(1-2\w)}{2\pi i\w^{N-k+1} (1-\w)^{N+k+1}},
\eeqa
where the last expression is a reformulation in terms of a loop
integral around $\w=0$. In a similar way $<P_k P_l>$ etc. can be
computed. We obtain
\beqa
	<P_k> &=& \frac{2N}{M_N} \left\{ \bin{2N-1}{N-k} - \bin{2N-1}{N-k-1} \right\}
	= \frac{2k}{M_N} \bin{2N}{N-k},
\\
	<P_k P_m> &=& \frac{2N}{M_N} \left\{ \bin{2N}{N-m} - \bin{2N}{N-k-m} \right\} , \; k \le m.
\eeqa
From these expressions, we can derive the following particular
averages, needed to compute the various energy moments over the
topology:
\beqa
\nonumber
	\left\langle\sum_k k P_k \right\rangle
	&=& \frac{N}{2}4^N\bin{2N}{N}^{-1}
	\approx \frac{N}{2} \sqrt{\pi N} \(1 + \frac{1}{8N} + \ldots\),
\\
\label{PP}
	\left\langle\sum_k k^2 P_k \right\rangle &=& N^2,
\\ \nonumber
	\left\langle\(\sum_k k P_k\)^2 \right\rangle &=& \frac{N^2(5N+1)}{6},
\eeqa
where the approximate form in the top equation is valid for large $N$.

%
\subsection{Full Energy Averages}
%

We are now ready to compute the final averages also over the topology.
Inserting the results of eqs. (\ref{PP}) into eqs. (\ref{aveE0}), we
obtain for the moments of the ground-state energy $E_0$ of a random problem:
\beqa
	<E_0> &=&
	\half 4^N \bin{2N}{N}^{-1} - \half
	\; \approx \;
	\half \sqrt{\pi N}
	- \half,
\\ \nonumber
	<E_0^2> &=&
	\frac{5N}{6}
	-\half 4^N \bin{2N}{N}^{-1}
	+ \half
	\; \approx \;
	\frac{5N}{6},
\\ \nonumber
	<E_0^2>_c & \approx & \frac{(10-3\pi)N}{12},
\eeqa
where $\langle ab\rangle_c = \langle ab\rangle-\langle a\rangle\langle
b\rangle$ is the connected moment; the approximate forms are valid for
large $N$.

Similarly, from eqs. (\ref{aveD}) we get for the moments of the
excitation energy $D$, for a random pairing in a random topology,
\beqa
	<D> &=& \frac{N+1}{2} - \half 4^N \bin{2N}{N}^{-1}
	\; \approx \;
	\frac{N}{2}
	- \half \sqrt{\pi N}
	+ \half,
\\
	<D^2> &=&
	\frac{N^2}{4}
	+ \frac{17N}{12}
	+ \frac{1}{3}
	- \frac{N+1}{2} 4^N \bin{2N}{N}^{-1}
\\ \nonumber
	&\approx &
	\frac{N^2}{4}
	- \frac{N}{2} \sqrt{\pi N}
	+ \frac{17N}{12},
\\
	<D^2>_c &\approx & \frac{(11-3\pi)N}{12}.
\eeqa

Averaging the combined moment, eq. (\ref{E0D}), over topology yields
\beqa
	<E_0 D> &=&
	\frac{N+2}{4} 4^N \bin{2N}{N}^{-1}
	- \frac{13N+5}{12}
	\; \approx \;
	\frac{N}{4} \sqrt{\pi N}
	- \frac{13N}{12},
\\
	<E_0 D>_c &=&
	- \frac{5N+1}{6}
	+ \frac{1}{4} 16^N \bin{2N}{N}^{-2}
\; \approx \;
	- \frac{(10 - 3\pi)N}{12}.
\eeqa

Combining this with the moments of $E_0$ and $D$, we get for the
moments of the total energy, $E=E0+D$, the following simple results
\beqa
	<E> &=& \frac{N}{2},
\\
	<E^2> &=& \frac{N^2}{4} + \frac{N}{12},
\\
	<E^2>_c &=& \frac{N}{12},
\eeqa
which can be understood by noting that a random pairing in a random
topology corresponds to a set of $N$ lines with random endpoints. Then
each line has a uniform length distribution between 0 and 1, and $E$
is their total length.

Computing the corresponding standard deviations, we have for a typical
random problem (and a random pairing, for $D$ and $E$) at large $N$,
\beqa
\label{fullE0}
	E_0 & \sim & \half \sqrt{\pi N}
	\; \pm \;
	\half \sqrt{\(\frac{10}{3} - \pi\) N},
\\
\label{fullD}
	D & \sim & \frac{N}{2} - \half \sqrt{\pi N}
	\; \pm \;
	\half \sqrt{\(\frac{11}{3} - \pi\) N},
\\
\label{fullE}
	E & \sim & \frac{N}{2}
	\; \pm \;
	\half \sqrt{\frac{N}{3}},
\eeqa
and we see that $E$ scales as $\sqrt{N}$, while $E$ and $D$ scale as
$N$, while the standard deviation scales as $\sqrt{N}$ in each case.

Of interest are also the correlations at large $N$, given to
order $N$ by
\beqa
\nonumber
	<E_0D>_c & \sim & -\frac{N}{12} ( 10 - 3 \pi ),
\\
	<E_0E>_c & \sim & 0,
\\ \nonumber
	<DE>_c & \sim & \frac{N}{12},
\eeqa
 indicating that $E$ and $E_0$ become uncorrelated for a random
pairing of a large random problem.

%
\subsection{Statistics for Individual Degeneracies}
%

An interesting but more difficult thing to compute is the average
fraction $\gamma_n = <g_n>/N!$ of pairings having a given excitation
energy $D = n$, in particular the ground-state fraction $\gamma_0$,
which gives the average probability of hitting a ground-state by
chance. Since $g_n$ is simply related to $a_n$, we will start by
considering
\beq
	\alpha_n = \frac{\bin{2N}{N}}{2N} <a_n>,
\eeq
in terms of which $\gamma_n$ can be expressed as
\beq
	\gamma_n = \frac{N!}{(2N-1)!} \sum_m (-)^m \bin{N+1}{m} \alpha_{n-m}.
\eeq
We then have
\beq
	\alpha_n = \frac{1}{2N} \sum_{[P]} m[P] \prod_k (n+k)^{P_k}
	= \sum_{[P]} \frac{1}{P_1} \delta_{N-\sum P_k}
	\prod_{k\ge 1}(n+k)^{P_k} \bin{P_k+P_{k+1}-1}{P_{k+1}}.
\eeq
A generating function for the $N$-dependence of $\alpha_n$ is then
\beq
	A_n(z) \equiv \sum_N z^N \alpha_n(N)
	= \sum_{[P]} \frac{1}{P_1}
	\prod_{k\ge 1}[z(n+k)]^{P_k} \bin{P_k+P_{k+1}-1}{P_{k+1}}.
\eeq
Again, starting the $P_k$ summation at a large $k_0$ and proceeding in
order of decreasing $k$, gives convergence of the full sum in the
limit $k_0 \to \infty$. Denoting by $\w_{k+n-1}^{P_k}$ the result of
summing above a certain $k$, we have by \eq{binsum},
\beq
\label{wrel}
	\w_{k-1} = \frac{kz}{1-w_k},
\eeq
and the final result is
\beq
	A_n(z) = \sum_{P_1} \frac{1}{P_1} \w_n^{P_1} = -\log\(1-\w_n(z)\).
\eeq
The recursion relation (\ref{wrel}) can now be linearized by assuming
$\w_k = p_k/q_k$, which can be solved e.g. by
\beqa
	q_{k-1} &=& q_k - p_k,
\\
	p_{k-1} &=& k z q_k,
\eeqa
which, upon eliminating $p_k$ gives
\beq
\label{recur}
	 k z q_k - q_{k-1} + q_{k-2} = 0.
\eeq

By partial integration, it is simple to prove that the following
sequence of integrals solves the recursion relation (\ref{recur}) for
$q_k$,
\beq
	q_k = \mbox{Im} \int_0^{\infty} \frac{(-iy)^k}{k!} \exp\(-z\frac{y^2}{2}+iy\) dy
	\;,\; k \ge 0,\;z > 0,
\eeq
which can be extended to negative $k$ by recursion.
The $q_k$ can be expanded in $z$ as
\beq
	q_k = \left \{
	\begin{array}{l}{}
	\sum_{m=0}^{\infty} \bin{2m+k}{2m}(2m-1)!!z^m \;,\; k \ge 0, \\
	\sum_{m=0}^{[-(k+1)/2]} \bin{-k-1}{2m}(2m-1)!!z^m \;,\; k < 0,
	\end{array}
	\right.
\eeq
where $[\,]$ denotes integer part. Note that for non-negative $k$ the
series is an asymptotic one, while for negative $k$ it is finite. In
particular, we have $q_{-1}=1$.
An independent solution to (\ref{recur}) is given by
\beq
	\hat{q}_k(z) = \frac{z^{-k}}{k!}q_{-k-1}(-z),
\eeq
but this solution is irrelevant, having the wrong large-$k$ behaviour.

In terms of $q_k$ we now have
\beqa
	1-\w_k &=& q_{k-1}/q_k,
\\
	A_n &=& \log\(q_n\) - \log\(q_{n-1}\),
\eeqa
which should then be expanded in powers of $z$ to yield $\alpha_n(N)$
as the coefficient of $z^N$. In particular, we have $A_0=\log(q_0)$,
where $q_0 = 1 + z + 3z^2 + 15z^3 + \ldots$.

In table \ref{g0tab}, results are displayed for the average degeneracy
of the two lowest states, based on an expansion of $A_0$ and $A_1$ in
powers of $z$.
\begin{table}[htb]
\small
\centering
\begin{tabular}{|r|r|r|r|r|r|r|r|}
\hline
\multicolumn{1}{|c} {$N$} &
\multicolumn{1}{|c} {$K_N=\half\bin{2N}{N}$} &
\multicolumn{1}{|c} {$K_N<g_0>$} &
\multicolumn{1}{|c} {$K_N<g_1>$} &
\multicolumn{1}{|c} {$<g_0>$} &
\multicolumn{1}{|c} {$<g_1>$} &
\multicolumn{1}{|c} {$\gamma_0$} &
\multicolumn{1}{|c|}{$\gamma_1$} \\
\hline
\hline
1 & 	1 & 	1 & 	0 & 	1.00000 & 	0.00000 & 	1.000000 & 	0.000000 \\
2 & 	3 & 	5 & 	1 & 	1.66667 & 	0.33333 & 	0.833333 & 	0.166667 \\
3 & 	10 & 	37 & 	22 & 	3.70000 & 	2.20000 & 	0.616667 & 	0.366667 \\
4 & 	35 & 	353 & 	411 & 	10.0857 & 	11.7429 & 	0.420238 & 	0.489286 \\
5 & 	126 & 	4081 & 	7676 & 	32.3889 & 	60.9206 & 	0.269907 & 	0.507672 \\
6 & 	462 & 	55205 & 	149741 & 	119.491 & 	324.115 & 	0.165960 & 	0.450159 \\
7 & 	1716 & 	854197 & 	3.09875$\times 10^6$ & 	497.784 & 	1805.80 & 	0.098767 & 	0.358294 \\
8 & 	6435 & 	1.4876$\times 10^7$ & 	6.84187$\times 10^7$ & 	2311.74 & 	10632.3 & 	0.057335 & 	0.263697 \\
9 & 	24310 & 	2.88019$\times 10^8$ & 	1.61447$\times 10^9$ & 	11847.7 & 	66411.7 & 	0.032649 & 	0.183013 \\
10 & 	92378 & 	6.13891$\times 10^9$ & 	4.07031$\times 10^{10}$ & 	66454.3 & 	440614. & 	0.018313 & 	0.121421 \\
11 & 	352716 & 	1.42882$\times 10^{11}$ & 	1.09496$\times 10^{12}$ & 	405092. & 	3.10436$\times 10^6$ & 	0.010148 & 	0.077771 \\
12 & 	1.35208$\times 10^6$ & 	3.60668$\times 10^{12}$ & 	3.13708$\times 10^{13}$ & 	2.66751$\times 10^6$ & 	2.32019$\times 10^7$ & 	0.005569 & 	0.048438 \\
13 & 	5.20030$\times 10^6$ & 	9.81584$\times 10^{13}$ & 	9.55147$\times 10^{14}$ & 	1.88755$\times 10^7$ & 	1.83672$\times 10^8$ & 	0.003031 & 	0.029496 \\
14 & 	2.00583$\times 10^7$ & 	2.86562$\times 10^{15}$ & 	3.08337$\times 10^{16}$ & 	1.42865$\times 10^8$ & 	1.5372$\times 10^9$ & 	0.001639 & 	0.017633 \\
\hline
\end{tabular}
\caption{Results for the average degeneracy of the lowest
 energy-states for various system sizes $N$. The second column gives
 an integer normalization factor $K_N$. Dividing
 the integers in the next two columns by $K_N$ yields the average
 number of ground-states $<g_0>$ and first excited states $<g_1>$,
 respectively.  Dividing these by $N!$ yields $\gamma_0$ and
 $\gamma_1$.}
\label{g0tab}
\end{table}
It is easy to check for small $N$, using table \ref{topol}, that
$g_m/\sum g_k$ for $m=0,1$, averaged over topologies with the proper
probabilities, indeed agrees with $\gamma_m$ of table \ref{g0tab},
obtained from the expansion of $A_k$.

The result for $\gamma_0$ strongly indicates an asymptotic behaviour
of $\gamma_0 \sim 2N2^{-N}$. This corresponds to an exponential
decrease with $N$ of the average probability for a random pairing to
hit a ground-state. However, the average number of ground-states grows
faster than exponentially: $<g_0> \sim 2N\;N!\;2^{-N}$.

This abundance of ground-states indicates that a corresponding
restricted problem might well have a solution with a locally minimal
waiting-time, if the restrictions are not too severe; this is used in
ref. \cite{long} to simplify the solution of a set of restricted crew
scheduling problems.

\newpage

By reducing the state-space of a restricted problem to the set of
ground-states of the corresponding unrestricted problem, the average
information gained at each airport is given by $\log \gamma_0$, which
for a large airport roughly yields $N \log 2$. Summing this over
several airports yields a total information gain scaling as $N_f \log
2$, with $N_f$ the total number of flights. Partly, this is due to a
grouping of flights, which contributes an average information gain of
$N_a \log 2$, with $N_a$ the number of airports.

%
\section{Conclusions}
%

We have performed a statistical analysis of an ensemble of random
unrestricted local crew scheduling problems, formulated in terms of
mapping arrivals onto departures at a single airport so as to minimize
waiting-time.

For the ground-state energy $E_0$ of a large random problem, we find
that both the average and the fluctuations scale like $\sqrt{N}$.  For
a random pairing of such a problem, on the other hand, the excitation
$D$ and the total energy $E = E_0 + D$ both grow linearly with $N$,
with fluctuations scaling like $\sqrt{N}$.

The individual degeneracies of the lowest energy states for random
problems are such that the average probability for hitting an optimal
pairing by chance decreases like $2N2^{-N}$ for large $N$. Since the
total number of pairings grows like $N!$, the average number of
ground-states grows very fast with system size.

The results and the methods of analysis described in this paper are
useful when designing efficient algorithms for the crew scheduling
problems with global restrictions, by providing means for estimating
the difficulty of a given problem, and for understanding and
simplifying its structure.

The optimal crew waiting-time for a restricted problem is bounded from
below by the ground-state energy of the corresponding unrestricted
problem, which is useful for gauging algorithmic performance for
problems of realistic size. Due to a faster than exponential growth of
the number of ground-states with problem size, this bound is often
saturated. This can be used to simplify a restricted problem: By
insisting on the local ground-state energy, airports can be split into
several parts. For a large random problem, this results on the average
in a reduction of state-space size by a factor of two for each airport.

Some of the calculations in this paper are based on novel methods of a
general nature, that may have applications also in other contexts with
a similar topological structure.

\end{document}